\documentstyle{article}

\title{\bf Properties of the hard pomeron with a running coupling constant
and the high-energy scattering}
\author{Mikhail Braun $^{a,b)}$\thanks{Visiting Professor IBERDROLA.
 Permanent address: Dep. High-Energy Physics, University of St. Petersburg,
198904 St.Petersburg, Russia}
, Gian Paolo Vacca  $^{a,b)}$ and Giovanni
Venturi$^{b)}$  \\
$^{a)}$Department of Particles, University of Santiago de Compostela\\
$^{b)}$Department of Physics, University of Bologna\\
$^{b)}$Istituto Nazionale di Fisica Nucleare - Sezione di Bologna.}
\date{May 1996}
\def\beq{\begin{equation}}
\def\eeq{\end{equation}}
\def\bea{\begin{eqnarray}}
\def\eea{\end{eqnarray}}
\def\noi{\noindent}
\def\oq{\omega(q)}
\def\oa{\omega(q_{1})}
\def\ob{\omega(q_{2})}
\def\eq{\eta (q)}
\def\ea{\eta (q_{1})}
\def\eb{\eta (q_{2})}
\def\ec{\eta (q'_{1})}
\def\ed{\eta (q'_{2})}
\def\fq{f (q)}
\def\fa{f (q_{1})}
\def\fb{f (q_{2})}
\def\fc{f (q'_{1})}
\def\fd{f (q'_{2})}
\def\ql{(q \cdot l)}
\def\qlp{(q \cdot l')}
\begin{document}
\maketitle
\medskip
\noi{\bf Abstract.}
The equation for the hard pomeron with a running coupling introduced on the
basis of the bootstrap requirement is solved numerically. Two supercritical
pomerons are found with the intercept minus 1 of the leading one of the
order 0.35--0.5 and that of the subleading one half as large. The
contribution of multipomeron exchanges is found to be essential only at
extremely high energies of the order of 100 $TeV$. Comparison of the
cross-sections and structure functions to the present experimental data seem
to indicate that the asymptotical regime has not yet been achieved.\vspace {4
cm}

{\Large\bf US-FT/23-96}

\newpage
\section{ Introduction.}
In  recently published papers [ 1,2 ] one of the authors has
 proposed a method to include the running
coupling constant into the dynamics of reggeized gluons based on the so-called
"bootstrap condition"[ 3 ]. This is a relation between the reggeized gluon
trajectory
\beq
\oq=-\alpha_{s}N_{c}q^{2}\int
\frac{d^{2}q_{1}}{(2\pi)^{2}}{q_{1}^{2}q_{2}^{2}},\ \
q=q_{1}+q_{2}
\eeq
and the gluon pair interaction given by the BFKL kernel [ 4 ]
\beq
K_{q}(q_{1},q'_{1})=-2T_{1}T_{2}\alpha_{s}
\left((\frac{q_{1}^{2}}{{q'_{1}}^{2}}+\frac{q_{2}^{2}}{{q'_{2}}^{2}})
\frac{1}{(q_{1}-q'_{1})^{2}}-\frac{q^{2}}{{q'_{1}}^{2}{q'_{2}}^{2}}\right)
\eeq
where $T$ is the gluon colour vector and $q_{1}+q_{2}=q'_{1}+q'_{2}=q$.
For the gluon channel $T_{1}T_{2}=-N_{c}/2$ and integrating the kernel one
obtains the bootstrap relation [ 3 ]
\beq
\int (d^{2}q'_{1}/(2\pi)^{2})K_{q}^{gluon}(q,q_{1},q'_{1})=
\oq-\oa-\ob
\eeq
This relation guarantees that in the $t$ channel with the gluon colour
quantum number the two gluon system has the reggeized gluon as its state.
As a result the production amplitudes in the one-reggeized-gluon-exchange
approximation, which serve as an input in the BFKL theory,
 satisfy unitarity in the leading order [ 5 ]. Thereby the whole scheme
becomes self-consistent: otherwise one should add to the input amplitudes
corrections following from the unitarity. Thus the bootstrap is a crucial
element for the reggeization of the gluon and  for the theory of the
reggeized gluons as a whole. Therefore the only way to introduce the running
coupling constant in a manner compatible with the gluon reggeization is to
preserve the bootstrap. This was the motivation of the papers [ 1,2 ]

Technically this can be achieved if one notices that (3) remains valid if
both in the gluon trajectory and interaction each of the momenta squared
is substituted by an arbitrary function of it. Indeed, if one takes for
the trajectory
\beq
\oq=-(N_{c}/2)\eq\int
\frac{d^{2}q_{1}}{(2\pi)^{2}}\frac{1}{{\ea\eb}},\ \
q=q_{1}+q_{2}
\eeq
and for the interaction kernel
\beq
K_{q}(q_{1},q'_{1})=-T_{1}T_{2}
\left( (\frac{\ea}{\ec}+\frac{\eb}{\ed})
\frac{1}{\eta(q_{1}-q'_{1})}-\frac{\eq}{\ec\ed}\right)
\eeq
then (3) will obviously be satisfied as well. The fixed coupling BFKL theory
corresponds to a particular choice
\beq
\eq=q^{2}/(2\alpha_{s})
\eeq

The idea of [1,2] was to change $\eq$ so that it  correspond to a running
rather a fixed coupling.  For the running coupling some conclusions about the
form of $\eq$ can be made considering the vacuum channel equation in the
limiting case of very large $q$. Then
putting (4) and (5) into the equation and assuming that $\eq$ grows with $q$,
in the leading approximation one gets an evolution equation
\beq
\partial \phi(q^{2},x)/\partial\ln 1/x=
(N_{c}/2\pi)\int^{q^{2}}\frac{dq_{1}^{2}}{\ea}\phi(q_{1}^{2},x)
\eeq
Comparison with the GLAP evolution equation in the leading order in $\ln
1/x$ (that is, in the double log approximation) allows to find the
asymptotic form of $\eq$:
\beq
\eq\simeq q^{2}/(2\alpha_{s}(q^{2})),\ \ q\rightarrow\infty
\eeq
It evidently differs from the fixed coupling case by changing the
fixed coupling constant $\alpha_{s}$ to a running one $\alpha_{s}(q^{2})$.
As a result, with a running coupling, both the gluon
trajectory and its interaction have to be changed simultaneously in an
interrelated manner, so that the resulting equation is different from the
BFKL one already in the leading order.

The behaviour of $\eq$ at small $q$, comparable or even smaller than the QCD
parameter $\Lambda$, cannot be established from any theoretical calculation,
since this domain is nonperturbative. In [ 1,2 ] these  confinement effects
were parametrizeed by an
effective "gluon mass" $m$, choosing $\eq$ in the form
\beq \eq=(b/2\pi)(q^{2}+m^{2})\ln ((q^{2}+m^{2})/\Lambda^{2})
\eeq
with $b=(1/4)(11-(2/3)N_{F})$ and $m\geq\Lambda$, which agrees with (8) for
large $q$ and remains finite up to $q=0$.

A preliminary study of the properties of the pomeron with $\eq$ given by (9)
was performed by the variational technique in [ 1,2 ]. It was found that
the intercept depended on the ratio $m/\Lambda$ quite weakly: as $m/\Lambda$
changes from 1.5 to 5.0 the intercept (minus one)  $\Delta$ falls from 0.4
to 0.25. On the other hand, the slope depends on this ratio very strongly.
This allows to fix the ratio $m/\Lambda$ to values in the interval
3.0$\div$4.0.

This variational study, although very simple, cannot however give values for
the intercept and especially for the slope with some precision.
 Still less can be found by this
method about the properties of the pomeron wave function essential for the
high-energy behaviour of the physical amplitudes [ 6 ]. Finally, one does not
receive any knowledge about the existence of other solutions with a
positive intercept. All these reasons give us a motivation to undertake a
numerical study of the two-gluon vacuum channel equation with the gluon
tralectory and interaction given by (4) and (5) respectively and the function
$\eq$ satisying (8). Its concrete form has been chosen to be slightly more
general than (9):
\beq \eq=(b/2\pi)f(q)
\eeq
where
\beq
f(q)=(q^{2}+m_{1}^{2})\ln ((q^{2}+m^{2})/\Lambda^{2})
\eeq
It allows for the freezing of the coupling and the confinement proper to occur
at somewhat different scales ($m$ and $m_{1}$ respectively). However, on
physical grounds, one feels that they should be of the same order.

In sections 2 and 3 we present the basic equations in the form suitable for
numerical analysis for the cases $q=0$ (forward scattering) and $q\neq 0$.
In Section 4 we describe the method of the solution and present the
numerical results for the intercept, slope and the wave function at $q=0$.
The results for the intercept and slope, on the whole, agree with those
found in [ 1,2 ] by the variational approach. An interesting new result is
the existence of a second pomeron with the intercept, roughly speaking,
two time less than for the leading one, but still positive.
In Section 5 these results are applied to study the asymptotical behaviour
of the cross-section for the $\gamma^{\ast}\gamma$ scattering.
Our conclusions are presented in Section 6.

\section{Basic equations. Pomeron at $q=0$}
We consider the physical case $N_{c}=3$. The units are chosen to have
$\Lambda=1$. In relating to observable quantities we take $\Lambda=0.2\ GeV$.

The pomeron equation is the eigenvalue equation
\beq
(-\oa-\ob)\phi(q_{1})+\int(d^{2}/(2\pi)^{2})
K_{q}^{vac}(q_{1},q'_{1})\phi(q'_{1})=
E(q)\phi(q_{1})
\eeq
where the "energy" eigenvalue  $E(q)$ is related to the pomeron trajectory via
\beq
\alpha(q)=1-E(q)=1+\Delta-\alpha' q^{2}
\eeq
The last equation, valid for small, $q$ defines the intercept $\Delta$ and the
slope $\alpha'$. In (12) the trajectories $\omega$ and the kernel $K^{vac}$ are
given by the Eqs. (4) and (5) with $T_{1}T_{2}=-3$ and
the function $\eta$ given by (10) and (11).
To symmetrize the kernel we pass to the function
\beq
\psi(q_{1})=\phi(q_{1})/\sqrt{\ea\eb}
\eeq
We also take out the common numerical factor
$6/((11-2/3N_{F})\pi)$ and express all terms via the
function $\fq$ defined by (11). Then the  equation for $\psi$ takes the
form
\beq
A_{q}(q_{1})\psi(q_{1})+\int d^{2}q'_{1}L_{q}(q_{1},q'_{1})\psi(q'_{1})=
\epsilon(q)\psi(q_{1})
\eeq
Here the "kinetic energy" is
\beq
A_{q}(q_{1})=(1/2) \int\frac{d^{2}q'_{1}\fa}{\fc f(q_{1}-q'_{1})}+
(1/2)\int\frac{d^{2}q'_{2}\fb}{\fd f(q_{2}-q'_{2})}
\eeq
The interaction kernel consists of two parts, a quasilocal  and a
separable ones:
\beq
L=L^{(ql)}+L^{(sep)}
\eeq
They are given by 
\beq
L^{(ql)}_{q}(q_{1},q'_{1})=
-\sqrt{\frac{\fa}{\fb}}\frac{1}{f(q_{1}-q'_{1})}\sqrt{\frac{\fd}{\fc}}-
\sqrt{\frac{\fb}{\fa}}\frac{1}{f(q_{2}-q'_{2})}\sqrt{\frac{\fc}{\fd}})
\eeq
and
\beq
L^{(sep)}_{q}(q_{1},q'_{1})=\frac{\fq}{\sqrt{\fa\fb\fc\fd}}
\eeq
Both parts are evidently symmetric in $q_{1}$ and $q'_{1}$.
The scaled energy $\epsilon$ is related to the initial one by
\beq
E=\frac{6}{\pi (11-(2/3)N_{F})}\epsilon
\eeq

Eq. (15) simplifies in the case when the total momentum of the two
gluons is equal to zero. With $q=0$ the two parts of the kinetic energies
become equal and the square roots in (18) turn to unity. So at $q=0$ the
equation retains its form (15) with
\beq
A_{0}(q_{1})=\int\frac{d^{2}q'_{1}\fa}{\fc f(q_{1}-q'_{1})}
\eeq
and the interaction given by (17) where now
\beq
L^{(ql)}_{0}(q_{1},q'_{1})=
-\frac{2}{f(q_{1}-q'_{1})}
\eeq
has really become local and
\beq
L^{(sep)}_{0}(q_{1},q'_{1})=\frac{f(0)}{\fa\fc}
\eeq
This is the equation which we shall solve numerically.

To make it one-dimensional we introduce the angular momentum of the gluons
$n$ and choose the solution in the form
\beq
\psi(q)=\psi_{n}(q^{2})\exp in\phi
\eeq
where $\phi$ is the azimuthal angle. Integrating over it in the
equation, we obtain an one-dimensional integral equation for the radial
function $\psi_{n}(q^{2})$:
\beq
A_{0}(q)\psi_{n}(q^{2})+\int dq_{1}^{2}
 L_{n}(q^{2},q_{1}^{2})\psi(q_{1}^{2})=
\epsilon\psi_{n}(q^{2})
\eeq
with the kernel  now given by
\beq
L_{n}(q^{2},q_{1}^{2})=-B_{n}(q^{2},q_{1}^{2})
+\delta_{n0}\pi\frac{f(0)}{\fq\fa}
\eeq
where
\beq
B_{n}(q^{2},q_{1}^{2})=\int_{0}^{2\pi} d\phi
\frac{\cos n\phi}{f(q^{2}+q_{1}^{2}-2qq_{1}\cos
\phi)}
\eeq
Note that $A_{0}$ can be expressed via $B_{0}$:
\beq
A_{0}(q)=(1/2)\int dq_{1}^{2}B_{0}(q^{2},q_{1}^{2})\frac{\fq}{\fa}
\eeq

Evidently Eq. (25) is very similar to  a two-dimensional
Shroedinger equation with an attractive interaction provided by the local
term and a positive kinetic energy described by $A$, which however grows
very slowly at high momenta (as $\ln\ln q$ according to  [ 1 ]).
Evidently the attraction becomes smaller with growing $n$. So we expect to
find negative energies, corresponding to intercepts larger than unity, only
for small $n$. Remember that for the BFKL pomeron only the isotropic state
with $n=0$ has a negative energy.  Our calculations reveal that  the
introduction of the running coupling  does not change this situation:
states with $|n|>0$ all have positive energies. So in the following
we consider the case $n=0$.

\section{Pomeron at $q\neq 0$: the slope}
With $q\neq 0$ the pomeron equation becomes essentially two dimensional.
Rather than to attempt to solve it numerically at all $q$ we limit ourselves
to small values of $q$ and determine not the whole trajectory $\alpha(q)$
but only the intercept $\alpha'$ defined by (13). This can be done in a much
simpler manner using a perturbative approach. We present "the Hamiltonian"
in (15)
\beq
H_{q}=A_{q}+L_{q}
\eeq
in the form
\beq
H_{q}=H_{0}+W(q)
\eeq
and calculate analytically $W(q)$ up to terms of the second order in $q$.
Then for small $q$ the value of the energy $\epsilon(q)$ will be given
 by the standard perturbation formula
\beq
\epsilon(q)=\epsilon (0)+<W(q)>
\eeq
where $<\ >$ means taking the average with the wave function at $q=0$,
determined from the numerical solution of the equation discussed in the
previous section. Thus we evade solving the two-dimensional problem, but, of
course, cannot determine more than the slope. Fortunately it is practically
all we need to study the high-energy asymptotics (although, of course, the
knowledge of the trajectory as a whole might be of some interest).

In order to derive an expression for $W(q)$ we pass to the relative momenta
$l$ and $l'$ 
\beq
q_1=(1/2)q+l, \quad q_2=(1/2)q-l, \quad q'_1=(1/2)q+l',\quad q'_2=(1/2)q-l'
\eeq
 Up to the second order in $q$ we have
\beq
\fa = f(l) [ 1 + a_1 \ql + \frac{a_1}{4}q^2 + \frac{a_2}{2}
	\ql^2 ]
\eeq
where
\beq
a_1 = a_1(l) = \bigl[1+\ln(l^2+m^2)-\frac{m^2-m_1^2}{l^2+m^2}\bigr]
\frac{1}{f(l)} \eeq
\beq a_2 = a_2(l) =
\bigl[\frac{1}{l^2+m^2}+\frac{m^2-m_1^2}{(l^2+m^2)^2}\bigr]\frac{1}{f(l)}
\eeq
 The expansion for $\fb$ differs by changing the sign of $l$ (or of $q$);
for $\fc$ and $\fd$ it suffices to replace $l$ by $l'$ in the expressions for
$\fa$ and $\fb$. We use also the notation $a'_1=a_1(l')$ and $a'_2=a_2(l')$.
We also need  the expansion for $f(q)$:
\beq
f(q)=f(0)(1+ a_3 q^2 ) + O(q^4)
\eeq
where
\beq
a_3 = \left(\frac{m_1^2}{m^2} + \ln m^2 \right) \frac{1}{f(0)}
\eeq

The perturbation $W(q)$, up to second order in $q$, can be expressed
via the introduced functions  $a_{1,2}$ and $a'_{1,2}$ and the constant
$a_{3}$. After some calculations we find a part of $W$ coming
from the kinetic term in Hamiltonian in the form
\bea
W_{1}(l)=
 &&\frac{1}{2}\int d^2 l' \frac{f(l)}{f(l')} \Bigl\{
 \bigl( \frac{1}{f(l-l')}+\frac{1}{f(-l-l')} \bigr)
 \Bigl[ -a'_1 \qlp -\frac{a'_1}{4} q^2 + \bigl( {a'}^2_1-\frac{a'_2}{2}\bigr)
 \qlp^2 \Bigr] + \nonumber \\
 &&\Bigl[ -\frac{a_1 a'_1 \ql \qlp}{f(l-l')} + \frac{a_1 a'_1 \ql \qlp}
 {f(-l -l')} \Bigr] \Bigr\}+ A_0(l) \bigl[\frac{a_1}{4} q^2+
\frac{a_2}{2}\ql^2\bigr] 
\eea
 The part of $W$ coming from the quasilocal part of the interaction
 can be written as
\beq
W_{2}(l,l')= 
  \frac{1}{2} \bigl[ a_1 ( q \cdot l) -a'_1 (q \cdot l') \bigr]^2
  L^{(ql)}_{0}(l,l')
\eeq
and the one coming from the separable part as
\beq
  W_{3}(l,l')=\bigl[
  \bigl( a_3 - \frac{a_1+a'_1}{4} \bigr) q^2 -\frac{1}{2} (a_2-a_1^2)
  (q \cdot l)^2  
  - \frac{1}{2} (a'_2-{a'}_1^{2}) (q \cdot l')^2 \bigr]
  L^{(sep)}_{0}(l,l')
\eeq

As mentioned, only isotropic solutions have the
intercept larger than one and are of interest. Then
the expression for $W(q)=\sum_{i=1,2,3} W_{i}$
has to be integrated over the azimuthal angles. 
Thus intergrated  values will be denoted
 $\hat{W}_i$, $i=1,2,3$.
Using (26)-(28), they can be conveniently expressed via the
kernel $B_{n}$ (eq. (27)):
  \beq
\frac {\hat{W}_1}{2 \pi q^2} =
  \frac{1}{4} \int d {l'}^2 \frac{f(l)}{f(l')} \bigl\{
  \bigl[ - \frac{a'_1}{2} + \bigl({a'}_1^2-\frac{a'_2}{2} \bigr) {l'}^2\bigr]
  B_0(l,l') -a_1 a'_1 l l' B_1(l,l') \Bigr\} + 
 \frac{A_0(l)}{4} \bigl( a_1+a_2 l^2 \bigr) \eeq\beq
\frac {\hat{W}_{2}}{2 \pi q^2} = - \frac{1}{2}
 \bigl( a_1^2 l^2 + {a'}_1^2 {l'}^2 \bigr)
  B_0(l,l') + a_1 a'_1 l l' B_1(l,l') \eeq\beq
\frac {\hat{W}_3}{2 \pi q^2} =
  \frac{f(0)}{f(l) f(l')} \Bigl[ 2 \pi \bigl(a_3 - \frac{a_1+a'_1}{4} \bigr)
     - \frac{1}{4} l^2 \bigl(a_2 - a_1^2 \bigr)
     - \frac{1}{4} {l'}^2 \bigl(a'_2 - {a'}_1^2 \bigr) \bigr] 
\eeq

The slope is given by the momentum average of the sum of these
 expressions, taken with a given isotropic wave function:
\beq
\alpha' = - \frac {(1/2)\int d l^2 d {l'}^2
   \psi(l) \psi(l') \bigl( \hat{W}_2 +\hat{W}_3 \bigr) +
	\int d l^2 \psi(l)^2 \hat{W}_1 }
	{   2\pi q^2 \int d l^2 \psi^2 (l)}
\eeq

\section{Numerical procedure and results}
Eq. (25) was first changed to the variable
$t=\ln q^{2}$
whereupon the wave function and the kernel transform according to
\beq \psi(q^{2})\rightarrow{\tilde\psi}(t)=q\psi(q^{2})\eeq
and
\beq L(q^{2},q_{1}^{2})\rightarrow{\tilde
L}(t,t_{1})=qq_{1}L(q^{2},q_{1}^{2})\eeq
Then the equation was reduced to a finite system of linear equations by
approximating the integral by a sum
\beq
\int_{-\infty}^{\infty}dt\,F(t)\simeq\sum_{i=1}^{n}w_{i}F(t_{i})
\eeq
with points $t_{i}$ and weights $w_{i}$ depending on the chosen
approximation scheme. The final equation is thus
\beq
\sum_{j=1}^{n}B_{ij}x_{j}=\epsilon x_{i},\ \ i=1,...n
\eeq
where
\beq
x_{i}=\sqrt{w_{i}}{\tilde \psi}(t_{i})
\eeq
and
\beq
B_{ij}=A(t_{i})\delta_{ij}+\sqrt{w_{i}w_{j}}{\tilde L}(t_{i},t_{j})
\eeq
A somewhat delicate part of procedure proved to be the integration over the
angle in (26), since at large values of $q^{2}$ and $q_{1}^{2}$ the
integrand becomes strongly peaked at $\phi=0$, so that one should take much
care to obtain a reasonable precision.

After determining the lowest eigenvalues $\epsilon$ and the corresponding
eigenvectors $x_{i}$ the wave function in the momentum space is directly
given by (45) and (49) at points $q^{2}=\exp t_{i}$. It should be normalized
according to
\beq
\int\frac{d^{2}q}{(2\pi)^{2}}|\psi(q)|^{2}=1
\eeq
Note that this wave function is a partially amputated one (see Eq. (14)).
The full (nonamputated) wave function is given by $\Phi(q)=\psi(q)/\eta(q)$.
It is this function that appears in the physical amplitudes.

The results for the lowest (and negative) eigenvalues of energy
for the case $n=0$ (isotropic pomeron) are
presented in Figs. 1 and 2. Actually energies with an opposite sign are
shown, which according to (13) give precisely the intercepts (minus one).
As mentioned, the QCD scale here and in the following is taken to be
$\Lambda=0.2\ GeV$. In Fig.1 the intercepts are shown for the case when the
two scales $m$ and $m_{1}$ in (11) are equal. Fig. 2 illustrates the
dependence of the intercepts on the ratio $m/m_{1}$.
The most interesting observation which follows from these figures at
once is that in all cases one observes two positive intercepts, which
correspond to two  different supercritical pomerons, the leading and
subleading ones. The intercept of the leading pomeron is found to be in
accordance with our earlier calculations, performed by the variational
method (it is slightly larger, which was to be expected). For physically
realistic values of $m$ and $m_{1}$ in the interval $0.5\div 1.0\  GeV$
it takes on values in the region of $0.5\div 0.3$ falling with the masses
$m$ and $m_{1}$. The same trend is seen for the intercepts of the subleading
pomeron, which lie in the interval $0.25\div 0.15$.

The slopes of the two found pomerons are shown in Fig. 3 as a function of
$m$ for the case $m=m_{1}$. The slopes depend very strongly on the value of
the regulator mass. The physically reasonable slopes for the domimant pomeron
of the order of $\alpha'\sim 0.25\ (GeV/c)^{-2}$ restrict the values of $m$ to
the region $0.7-0.9\ GeV$. So finally we choose 
\beq
m=0.82\ GeV\eeq
which leads to the following parameters of the leading (0) and subleading (1)
pomerons
\beq
\Delta_{0}=0.384,\ \ \alpha'_{0}=0.250\ (GeV/c)^{-2};\ \ 
\Delta_{1}=0.192,\ \ \alpha'_{1}=0.124\ (GeV/c)^{-2}
\eeq
In Fig. 4 we show the coordinate space wave functions $\Phi(r)$  of
these two pomerons. 

\section{Pomerons and the high-energy  scattering. Discussion}
To apply the found results to the actual physical processes one has to couple
the pomerons to the external sources corresponding to the colliding
particles. The only way to do it in a more or less rigorous manner is to
assume that both the projectile and target are highly virtual photons with
momenta $q$ and $p$ respectively, $-q^{2}=Q^{2}>>\Lambda^{2}$ and
$-p^{2}=P^{2}>>\Lambda^{2}$. Then the nonperturbative effects inside the
target and projectile can safely be neglected. As shown in [6] the scattering
amplitude in the high colour number limit then takes an eikonal form for fixed transverse dimensions of the
projectile and target and leads to a cross-section
\beq
\sigma=2\int d^{2}Rd^{2}rd^{2}r'\rho_{q}(r)\rho_{p}(r')
\left(1-\exp(-z(\nu,R,r,r'))\right)
\eeq
where
\beq
z(\nu,R,r,r')=(1/8)\int\frac{d^{2}qd^{2}q_{1}d^{2}q'_{1}}{(2\pi)^{6}}
G(\nu,q,q_{1},q'_{1})\exp iqR\prod_{i=1,2}(1-\exp iq_{i}r)(1- \exp iq'_{i}r')
\eeq
is essentially the Fourier transform of the (nonamputated) Green function of
Eq. (12), $G(\nu,q,q_{1},q'_{1})$, considered as a function of the energetic
variable $\nu=pq$ and with $q=q_{1}+q_{2}=q'_{1}+q'_{2}$. The functions
 $\rho_{q}$ and $\rho_{p}$ correspond to the colour densities of the
projectile and target photons, respectively. Their explicit form was found in
[7] for both transverse and longitudinal photons.

The found supercritical pomerons represent a part of the total pomeron
spectrum which contributes to the Green function a term dominating at high
energies \beq
G_{P}(\nu,q,q_{1},q'_{1})=\sum_{i=0,1}\nu^{\alpha_{i}(q)-1}
\Phi_{i}(q_{1},q_{2})
\Phi_{i}^{\ast}(q'_{1},q'_{2})
\eeq
where $\alpha_{i}$ and $\Phi_{i}$ are the trajectories and wave functions of
the leading (0) and subleading (1) pomerons. At high $\nu$ we can neglect the
dependence on the total momentum $q$ of the wave functions, taking them at
$q=0$, and approximate the trajectories according to (13). Then all the
quantities in (54) become determined, so that we can calculate the
cross-sections for both the transversal and longitudinal projectile photon 
and thus find the structure function of the virtual photon target. We have
taken for the latter a transversal photon with the lowest momentum
admissible of $P=1 \ GeV/c$. The resulting structure functions are presented
in Fig. 5 for the interval of small $x$ which we extended to extraordinary
small values to clearly see the unitarization effects.

To move closer to reality one has to consider hadronic target and
projectiles. The confinement effects make any rigorous treatment of
such a case impossible. Rather than to introduce arbitrary parameters (in
fact, functions) we extend the formula (54) to hadronic target and projectile
 subsituting the photonic colour densities by hadronic ones. For the latter
we take a Gaussian form and a normalization which follows from the comparison
to the electromagnetic densities with only the simplest quark diagrams taken
into account. In particular for the proton we take the Gaussian $\rho$, with
the observed electromagnetic proton radius and normalized to three. Such a
treatment, in all probability, somewhat underestimates the density, since it
does not include coupling to gluons.

The proton structure functions and the proton-proton total cross-sections
which follow from this approximation for the densities are shown in Figs. 6
and 7 respectively.
To see the unitarization effects we had again  to
consider extraordinary high values of $1/x$ and energies, well beyond our
present experimental possibilities.

In discussing these results, we have first to note that their overall
normalization is somewhat undetermined, since the exact scale at which $\nu$
enters into $\ln\nu$ factors is unknown. A second point to note is that
the subleading pomeron contribution is always very small: it amounts to a few
percent at smallest values of $1/x$ and $s$ considered and naturally gets
still smaller at higher $1/x$ or $s$.

As one observes from Figs. 5-7, the structure functions and cross-sections
monotonously rise with $1/x$, $s$ and $Q^{2}$. Studying the asymptotics of
the solutions of Eq. (12) at high $q$ and of Eq. (54) one can show that this
rise is logarithmic. In particular, the structure function of the virtual
photon rises as $\ln^{4}(1/x)$ and as $\ln^{\beta}(q^{2})$ with $\beta\sim
2.5$. The proton-proton cross-section eventually rise as $\ln^{2}s$, as
expected. Comparison to the Froissart bound (dash-dotted line in Fig. 7)
shows however that it remains far from being saturated.

The most interesting result that follows from Figs. 5-7 is that the
unitarization effects become visible only at exceedingly very small values of
$x$ or very large values of $s$, well outside the range of the present
experiment. They appear earlier at lower $Q^{2}$. Still at the smallest value
$Q=2\ GeV/c$ considered, the exchange of more than one pomeron achieves 
only 15\% of
the total for the  proton structure function at $x=10^{-10}$. Likewise the
relative contribution of many pomerons to the proton-proton cross-section
rises to 23\% only at $s\sim 10^{5} GeV$.

Comparing the calculated proton structure functions and the cross-sections
with the experimental results at highest $1/x$ and $s$ available we observe
that our results are essentially smaller than the observed ones.
Experimental value of $F_{2p}(Q^{2},x)$ at $Q^{2}=8.5\ (GeV/c)^{2}$ and
$x=0.000178$ is $1.19\pm 0.05\pm 0.16$ [8 ]. Our calculations only give a
value 0.17. The $\bar p p$ cross-section at $\sqrt{s}=1800\ GeV$ is around
$80\ mbn$ [9], whereas our result is $18.5\ mbn$. Of course, having in mind 
the uncertainties in the overall normalization and a very crude picture for
the pomeron coupling to the proton assumed, one cannot ascribe too much
importance to this fact. However one is tempted to explain this
underestimation of the experimental values by the simple reason that we are
 too far from the pure asymptotical regime yet and that other solutions of
Eq. (13)
different from the found supercritical pomerons and having their intercepts
around unity give the bulk of the contribution at present energies.
This may also explain the notorious discrepancy between a
high value of the hard pomeron intercept, of the order 0.35--0.5, and the
observed slow growth of the experimental cross-section, well described by the
"soft pomeron" with an intercept around 0.08. 

If this picture is correct then we may expect that with the growth of energy
the cross-sections will grow faster and faster, until at $\sqrt{s}\sim
10\ TeV$ they will become well described by a pure hard pomeron with
the intercept  0.35--0.5. This power growth will
continue until energies of an order $1000\ TeV$ when finally the
unitarity corrections set in to moderate the growth in accordance with the
Froissart bound.

\section{Conclusions}
The result of our study show that with the running coupling included the
pomeron equation posseses bound state solutions which have negative energy
and thus intercepts greater than unity. These solutions correspond to
supercritical pomerons in the old sense, that is, they represent simple
poles in the complex angular momentum plane. A new result is that two such
pomerons appear. However the subdominant pomeron does not seem to play any
role in describing the asymptotical behaviour of the amplitudes. The
intercepts of the pomerons depend weakly on the infrared regulator
parameter and stay in the region 0.35--0.5 for its physically reasonable
values. The introduction  of the running coupling and thus a scale  provides
for a notrivial slope for the pomeron, which is responsible for the
physically reasonable behaviour of the cross-sections at very high energies.

For realistic photonic cross-sections and with a rather crude approximation
for the hadronic ones unitarization effects begin to be felt at
extraordinary high energies, of the order $100-1000\ TeV$  (or equivalently
$x<10^{-10}-10^{-12}$). Until these energies a single pomeron exchange
remains a very good approximation to the asymptotic amplitude.

Comparison to the experimental cross-sections and structure functions at the
highest energy (lowest $x$) achieved seems to confirm the widespread opinion
that we are still rather far from the asymptotical regime and that other
states, different from the supercritical pomerons, give the dominant
contribution.

\section {Acknowledgments.}
The authors express their gratitude to Prof. C.Pajares for his attention and
helpful discussions.  
 M.A.B thanks the INFN for financial help during his stay
at Bologna University and IBERDROLA for financial support during his stay at
the University of Santiago de Compostela. G.P.Vacca thanks Profs. C.Pajares
and L.Miramontes for their hospitality during his stay at the University of
Santiago de Compostela.

\newpage
\section{References.}
\noi 1. M.A.Braun, Phys. Lett. {\bf B345} (1995) 155.\\
\noi 2. M.A.Braun, Phys. Lett. {\bf B348} (1995) 190.\\
\noi 3. L.N.Lipatov, Yad. Fiz. {\bf 23}(1976) 642.\\
\noi 4. V.S.Fadin, E.A.Kuraev and L.N.Lipatov,
 Phys. Lett. {\bf B60} (1975)50.\\
I.I.Balitsky and L.N.Lipatov, Sov.J.Nucl.Phys. {\bf 15} (1978) 438.\\
\noi 5. J.Bartels, Nucl. Phys. {\bf B151} (1979) 293.\\
\noi 6. M.A.Braun, Phys. Lett. {\bf B357} (1995) 138\\
\noi 7. N.N.Nikolaev and B.G.Zakharov, Z.Phys. {\bf C49} (1991) 607\\
\noi 8. H1 Collab., Nucl. Phys. {\bf B439} (1995) 471\\
\noi 9. See M.Block et al in {\it Proc. 24th Int. Symp. on Multiparticle
Dynamics}, Ed. A.Giovanini, World Scie. (1995).

\newpage
\section{Figure captions}
\noi Fig. 1 Pomeron intercepts as a function of the infrared regulator mass
$m=m_{1}$; $\Lambda=0.2\ GeV$.\\
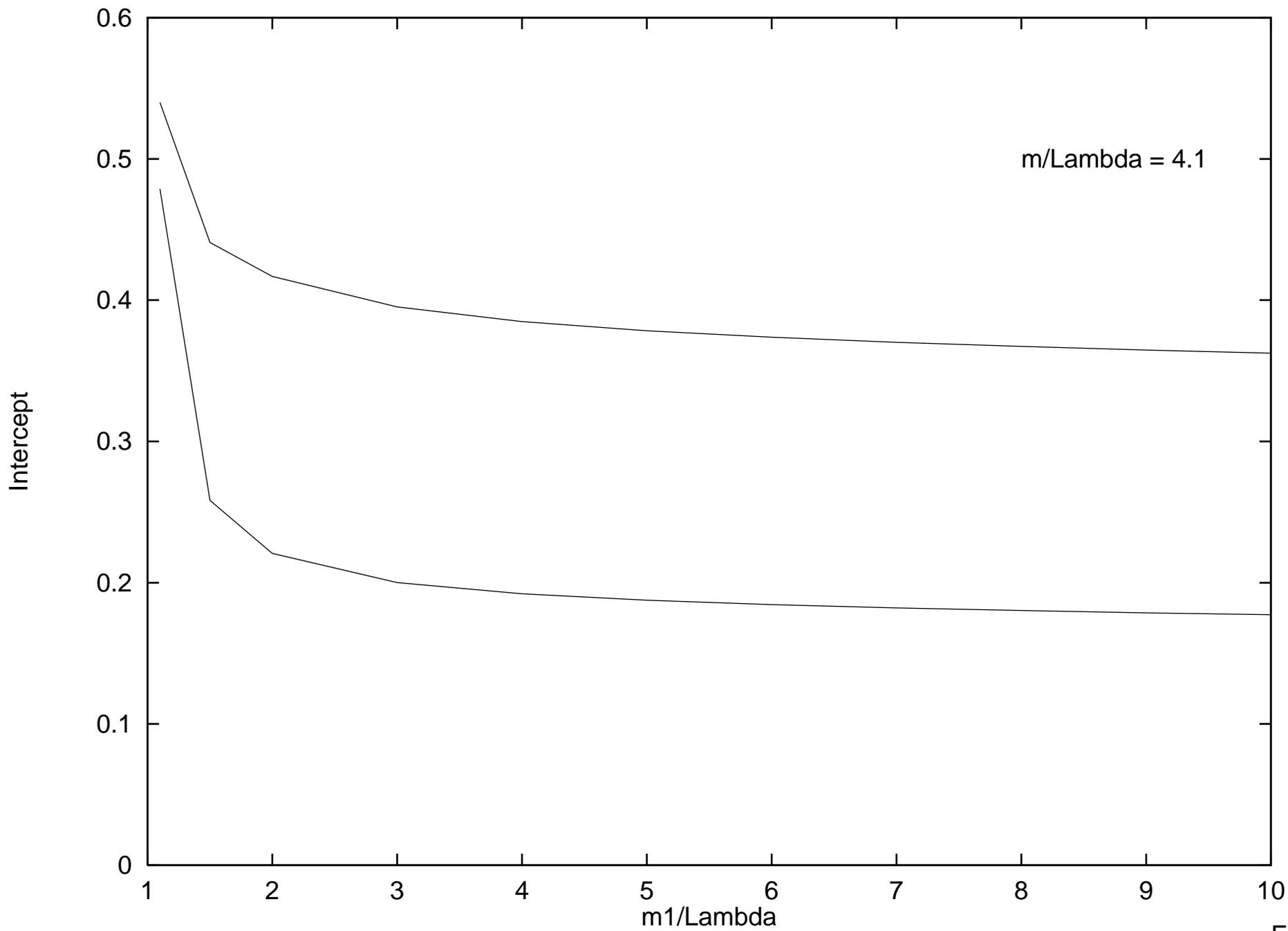
Fig. 2. Pomeron intercepts for different values of the confinement parameter
$m_{1}$ and the coupling freezing parameter $m$.\\
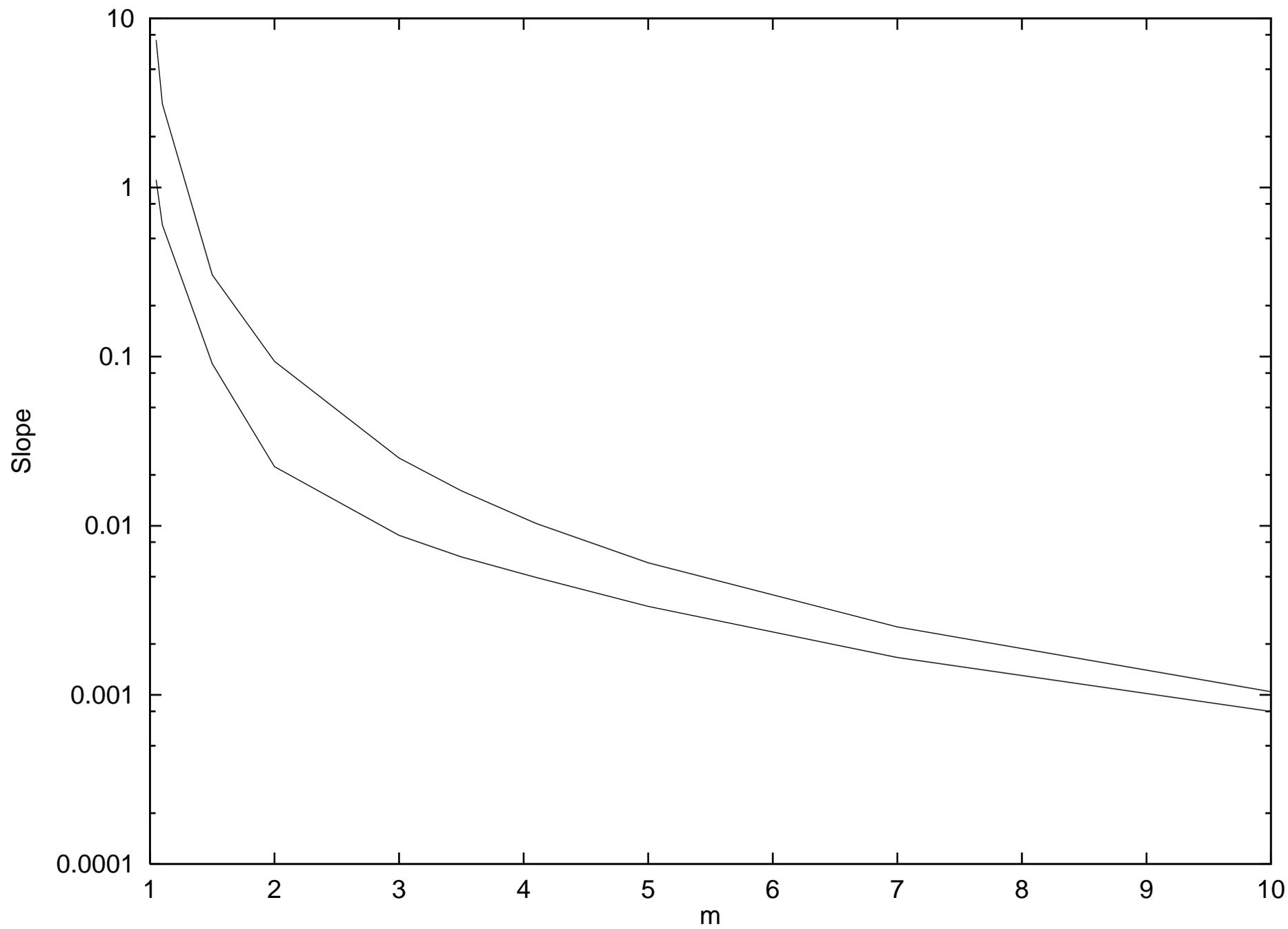
Fig. 3. Pomeron slopes as a function of the infrared regulator mass
$m=m_{1}$; $\Lambda=0.2\ GeV$.\\
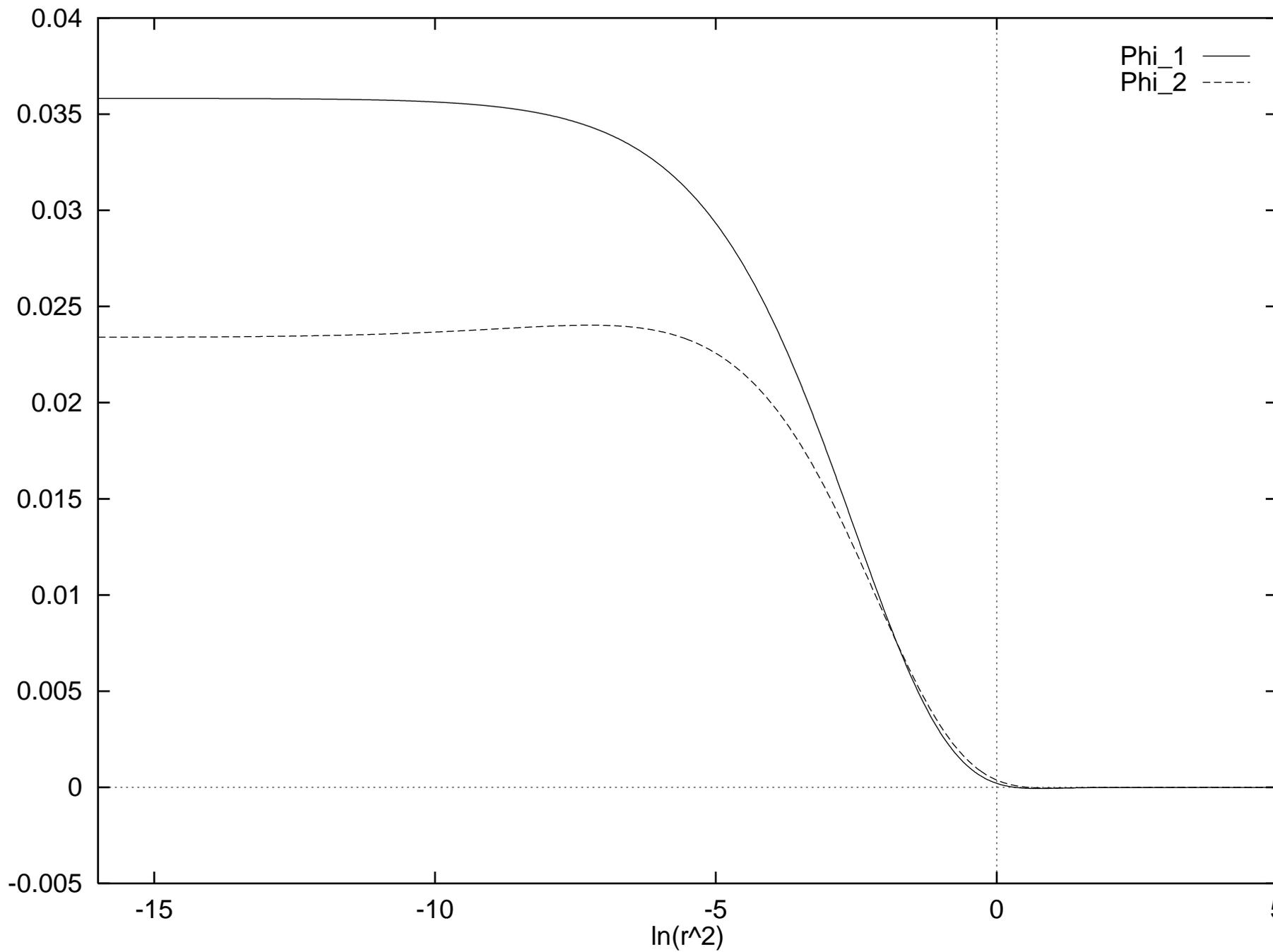
Fig. 4. Coordinate space wave functions for the leading  ($\Phi_{0}(r)$) and
subleading ($\Phi_{1}(r)$) pomerons. Both $r$ and $\Phi$ are in units
$1/\Lambda\simeq1\ fm$\\
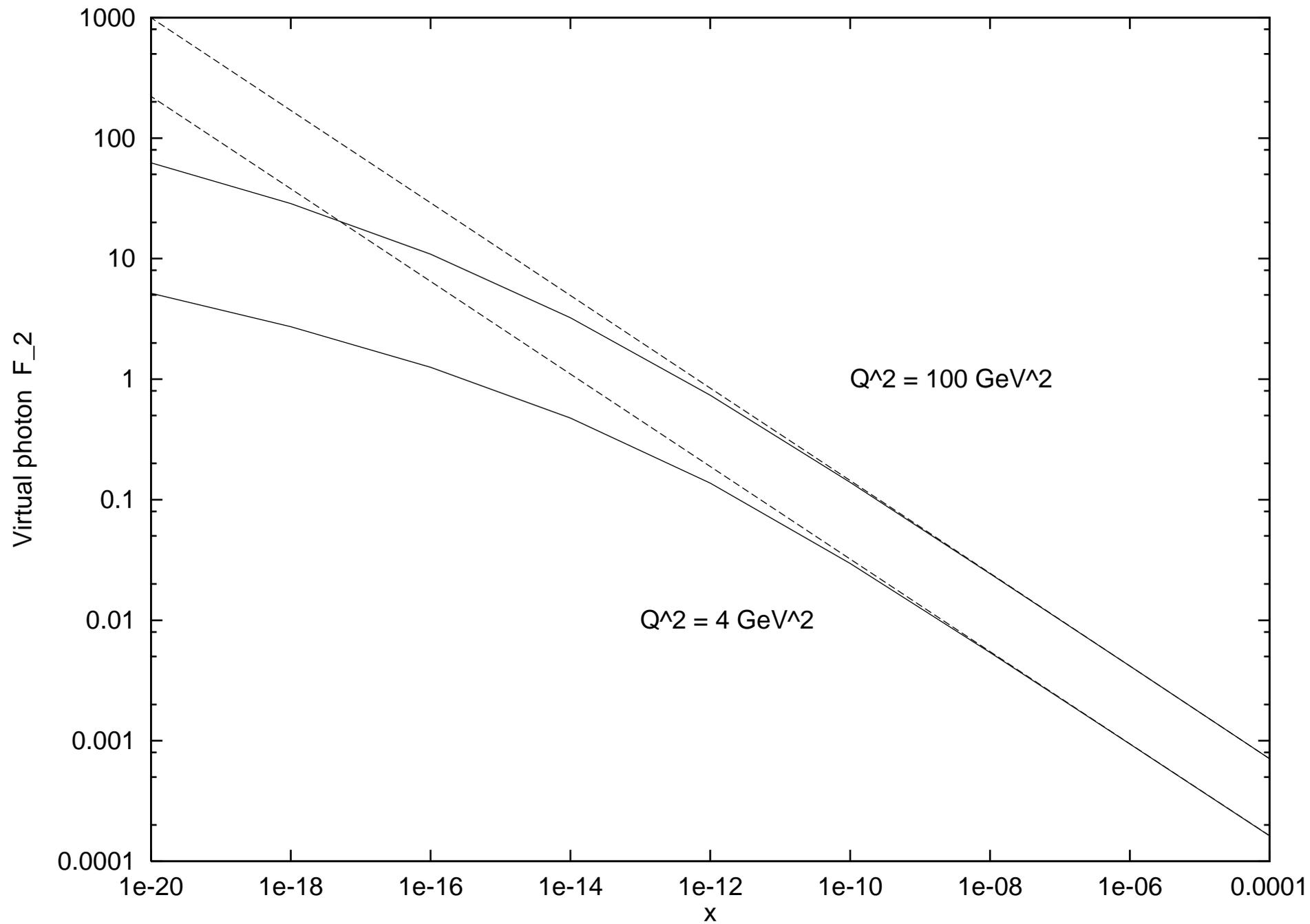
Fig. 5. Structure functions of a virtual photon ($P=1\ GeV/c$) at $Q^{2}=
4$ and $100\ (GeV/c)^{2}$ as a function of $x$ (solid curves). Dashed curves
show the contribution of a single pomeron exchange.\\
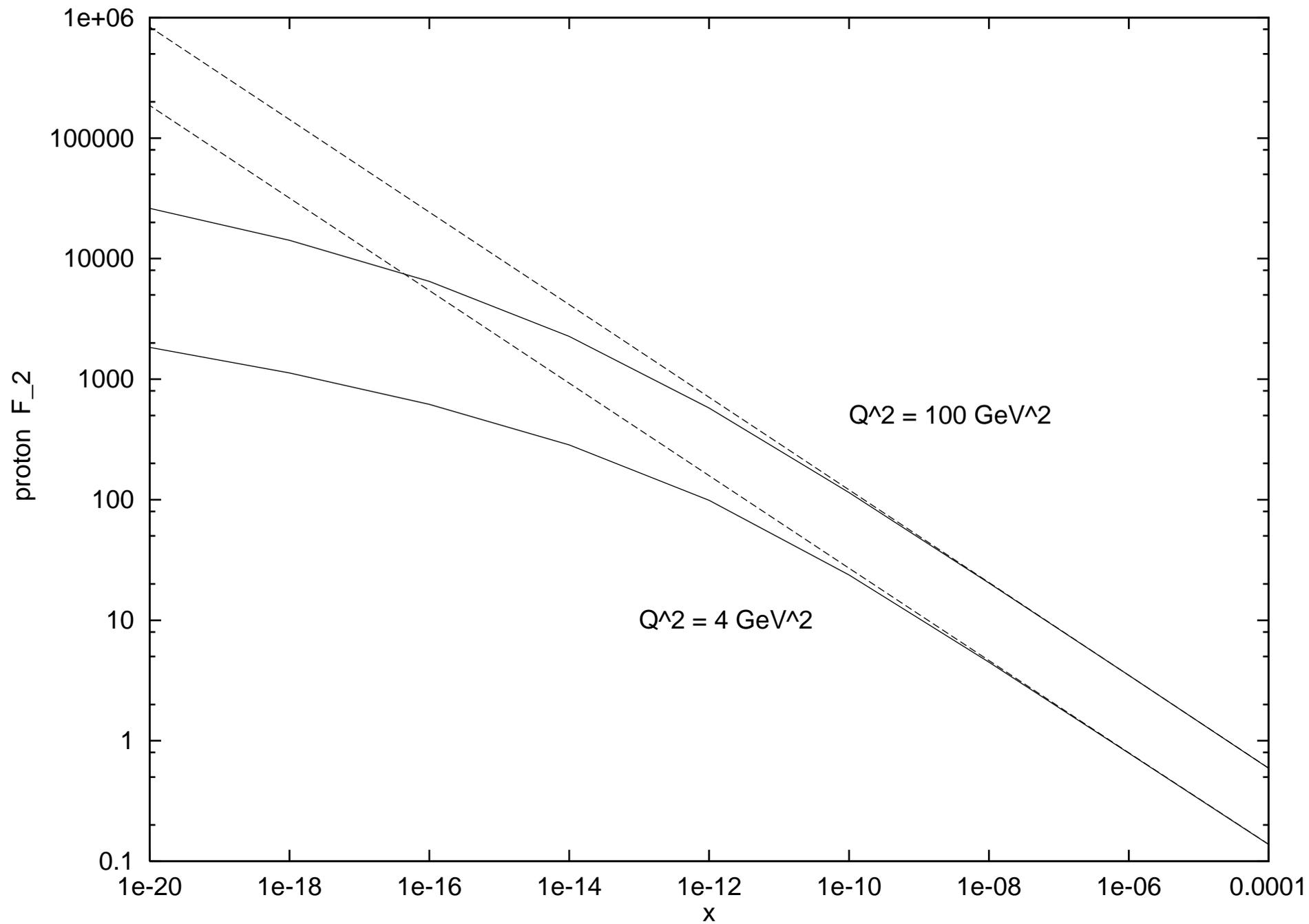
Fig. 6. Proton structure functions  at $Q^{2}=
4$ and $100\ (GeV/c)^{2}$ as a function of $x$ (solid curves). Dashed curves
show the contribution of a single pomeron exchange.\\
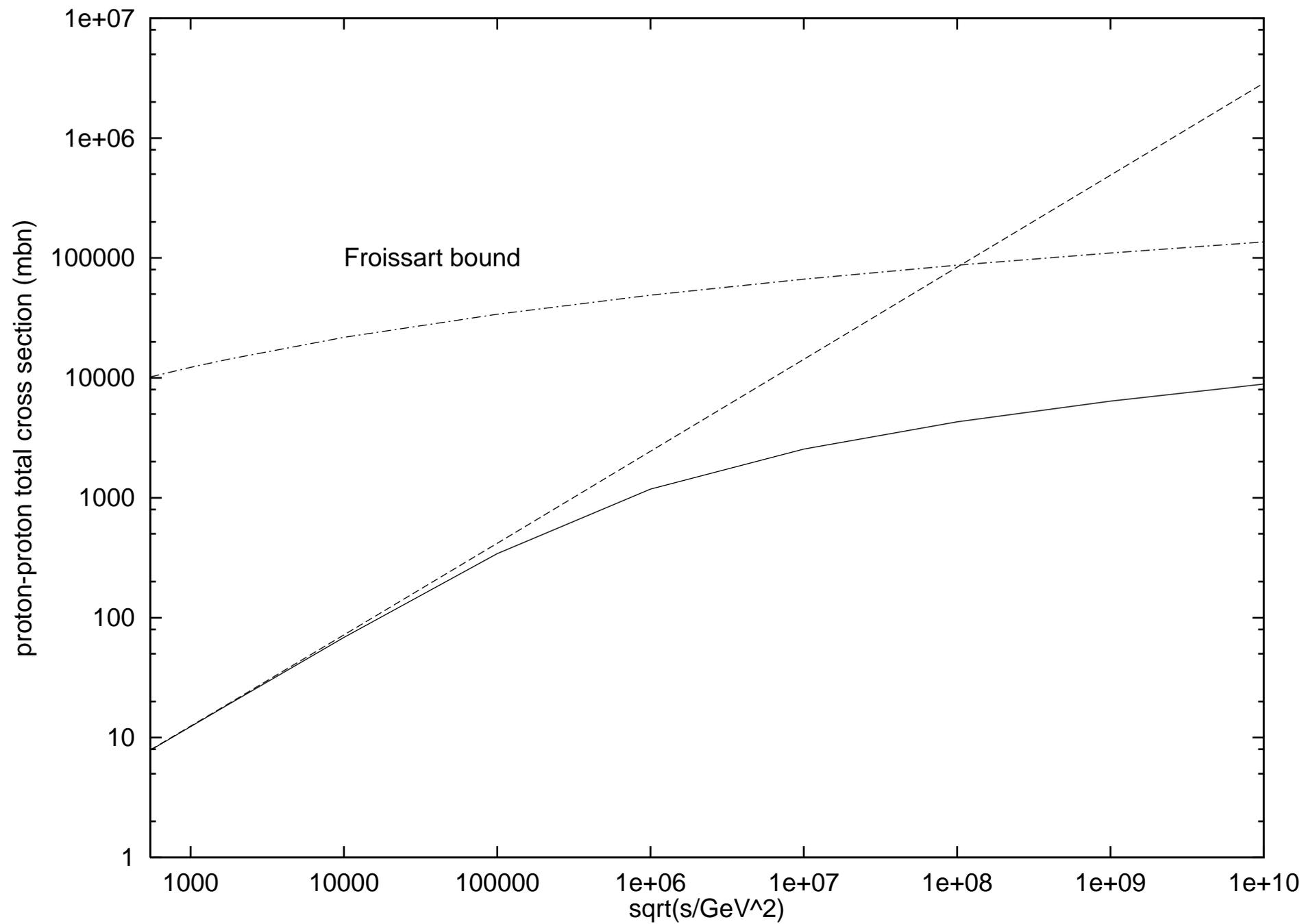
Fig. 7. Proton-proton total cross-sections
as a function of c.m. energy $\sqrt{s}$ (the solid curve). The dashed curve
shows the contribution of a single pomeron exchange. The dash-dotted curve
marks the Froissart bound.\\
 
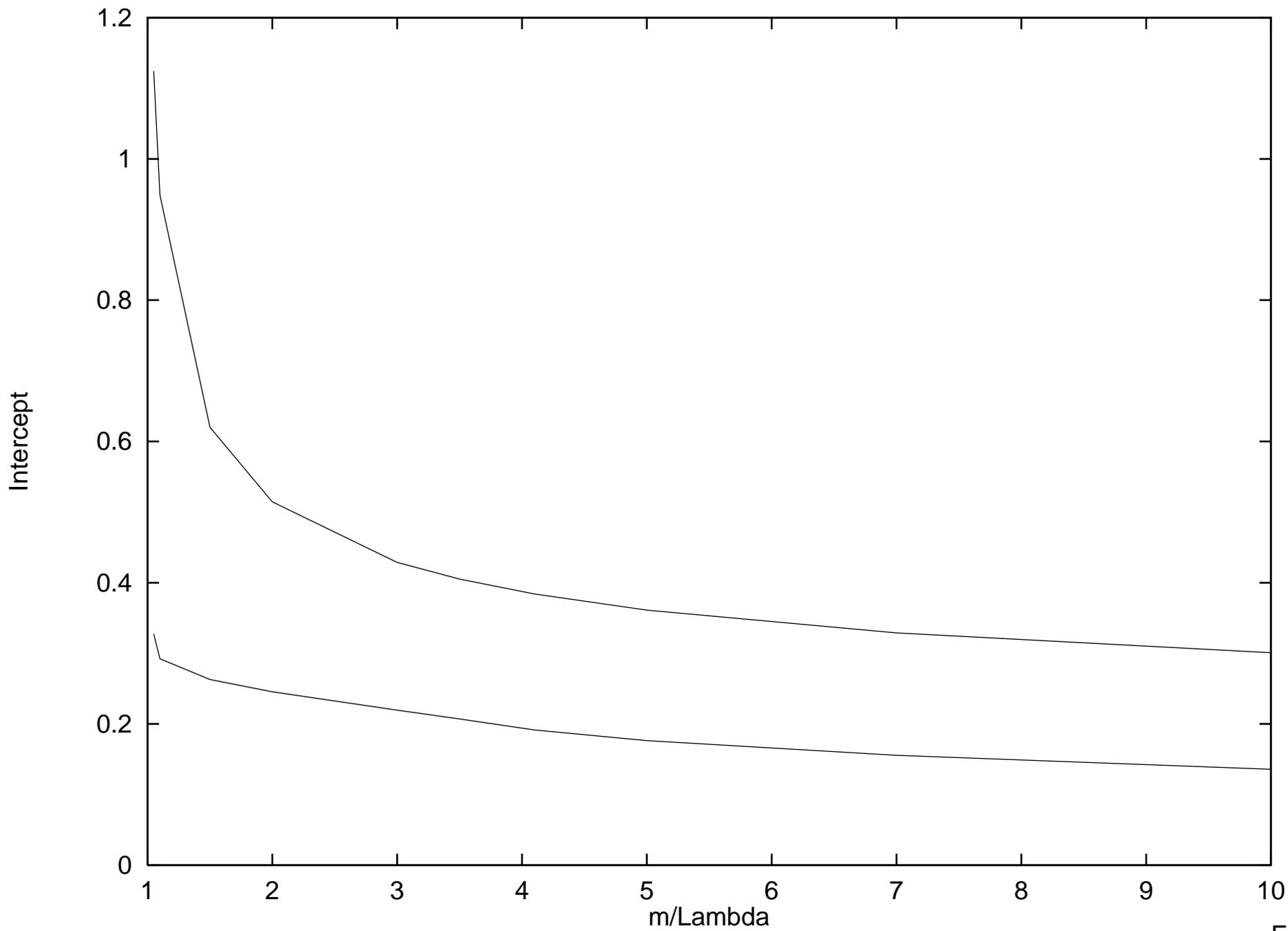
\end{document}